\documentclass[conference]{IEEEtran}
\IEEEoverridecommandlockouts
\usepackage{cite}
\usepackage{amsmath,amssymb,amsfonts}
\usepackage{algorithmic}
\usepackage{graphicx}
\usepackage{textcomp}
\usepackage{xcolor}
\usepackage{booktabs}
\usepackage{multirow}
\usepackage{orcidlink}
\def\BibTeX{{\rm B\kern-.05em{\sc i\kern-.025em b}\kern-.08em
    T\kern-.1667em\lower.7ex\hbox{E}\kern-.125emX}}
\begin{document}

\title{
Flatten Wisely: How Patch Order Shapes Mamba-Powered Vision for MRI Segmentation
\thanks{This work was supported by Qatar Research, Development, and Innovation (QRDI) Council under UREP Grant No.\ UREP31-077-2-025.}
\thanks{$^{*}$Equal contribution}
}

\author{
\IEEEauthorblockN{Osama Hardan$^{*}$}
\IEEEauthorblockA{\textit{Department of Computer Science and Engineering,} \\
\textit{Qatar University, Doha, Qatar}\\
oh2107304@qu.edu.qa \orcidlink{0009-0003-5703-6960}}
\and
\IEEEauthorblockN{Omar Elshenhabi$^{*}$}
\IEEEauthorblockA{\textit{Department of Computer Science and Engineering} \\
\textit{Qatar University, Doha, Qatar} \\
oe2106273@qu.edu.qa \orcidlink{0000-0002-7974-6061}}
\and
\IEEEauthorblockN{Tamer Khattab}
\IEEEauthorblockA{\textit{Department of Electrical Engineering} \\
\textit{Qatar University, Doha, Qatar.} \\
tkhattab@qu.edu.qa}

\and
\IEEEauthorblockN{Mohamed Mabrok}
\IEEEauthorblockA{\textit{Department of Mathematics and Statistics} \\
\textit{Qatar University, Doha, Qatar} \\
m.a.mabrok@qu.edu.qa}
}

\maketitle

\begin{abstract}
Vision Mamba models promise transformer-level performance at linear computational cost, but their reliance on serializing 2D images into 1D sequences introduces a critical, yet overlooked, design choice: the patch scan order. In medical imaging, where modalities like brain MRI contain strong anatomical priors, this choice is non-trivial. This paper presents the first systematic study of how scan order impacts MRI segmentation. We introduce \textbf{Multi-Scan 2D (MS2D)}, a parameter-free module for Mamba-based architectures that facilitates exploring diverse scan paths without additional computational cost. We conduct a large-scale benchmark of 21 scan strategies on three public datasets (BraTS 2020, ISLES 2022, LGG), covering over 70,000 slices. Our analysis shows conclusively that scan order is a statistically significant factor (Friedman test: $\chi^{2}_{20}=43.9, p=0.0016$), with performance varying by as much as 27 Dice points. Spatially contiguous paths---simple horizontal and vertical rasters---consistently outperform disjointed diagonal scans. We conclude that scan order is a powerful, cost-free hyperparameter, and provide an evidence-based shortlist of optimal paths to maximize the performance of Mamba models in medical imaging.
\end{abstract}


\section{Introduction}

Magnetic resonance imaging (MRI) is the primary modality for visualising soft-tissue pathology in the brain. Automated slice-level segmentation of tumours, oedema and ischaemic stroke lesions accelerates clinical workflows, supports longitudinal outcome studies and reduces the variability that plagues manual delineation. Convolutional neural networks (CNNs) such as U-Net have driven the last decade of progress by learning local texture cues efficiently \cite{ronneberger2015unet}. Despite continual refinements in depth, width and receptive-field manipulation, CNNs still extract context through cascades of limited-scope filters; the effective receptive field covers only a small fraction of the theoretical area \cite{luo2017understanding}, which hampers modelling of long-range dependencies critical for accurate neuro-anatomical segmentation.

Transformer-based architectures extend the receptive field to the entire image by applying self-attention across patch tokens. Although this mechanism has produced impressive gains in natural and medical vision, its memory footprint scales quadratically with the number of patches. For high-resolution brain MRI or multi-volume studies, quadratic scaling translates into prohibitive cost during training and inference, constraining adoption in clinical environments where GPU capacity and latency budgets are tight \cite{dosovitskiy2020image,kainz2024deep}.

State-space sequence models (SSMs) offer a compelling alternative. By framing sequence processing as a discretised linear time-invariant system, SSMs enable convolutional implementations that run in linear time and constant memory with respect to sequence length \cite{gu2021s4}. Mamba augments this framework with an input-dependent selective gate and a parallel-scan kernel that preserve global context while lowering overhead \cite{gu2023mamba}. Vision-adapted variants—including Vision Mamba and VMamba—embed Mamba blocks in hierarchical backbones, serialising the two-dimensional image into a one-dimensional patch sequence before processing \cite{zhu2024vision,liu2024vmamba}.

Serialisation introduces a design variable in state-space vision models: the \textbf{scan order}. Each scan defines how an \(H\times W\) grid of patches is traversed and therefore decides which spatial neighbours appear adjacent in the sequence. Adjacency in sequence space influences how easily the model can rebuild local structure; poorly chosen orders may force the network to expend capacity restoring spatial coherence rather than modelling semantics. Existing Vision Mamba studies adopt fixed heuristics—usually four orthogonal directions or a bidirectional raster inherited from natural-image practice \cite{zhu2024vision,liu2024vmamba,yang2024plainmamba}. Zhu \textit{et al.} systematically ablated twenty-two orders on aerial imagery and concluded that a single left-to-right scan sufficed for land-cover segmentation \cite{zhu2024rethinking}. Whether this conclusion generalises to brain MRI is unknown. Neuro-anatomical data possess strong directional regularities: white-matter tracts align along preferred axes, and cortical gyri and vascular territories introduce patterned anisotropy \cite{alexander2007diffusion,lee2009anisotropic}. These priors suggest that patch order may influence performance more strongly in brain MRI than in the less structured layout of remote-sensing scenes. A non-parametric Friedman test across LGG, BraTS 2020 and ISLES 2022 rejects the null hypothesis that all scan orders perform equally (\(\chi^{2}_{20}=43.9,\; p=0.0016\)), confirming that scan direction matters for brain-MRI segmentation.

To investigate this effect we extend VMamba with \textit{Multi-Scan-2D} (MS2D), a module that offers twelve primitive scan paths instead of the default four while preserving parameter count and runtime. MS2D feeds four parallel Mamba streams inside each residual unit; users select one, two or four unique directions, and the module duplicates them as needed so that exactly four streams are processed, keeping computational load constant. Leveraging MS2D we construct a benchmark of twenty-one scan strategies—twelve unidirectional scans, six bidirectional pairs and three balanced four-direction mixes—and test them on three public brain-MRI datasets: LGG for low-grade glioma, BraTS 2020 for heterogeneous high-grade tumours and ISLES 2022 for acute ischaemic stroke. Together these datasets contain more than seventy thousand slices.

Our contributions are threefold.  
\begin{enumerate}
  \item \textbf{Architecture}: We introduce MS2D, a plug-in for VMamba that expands the scan library to twelve directions without adding parameters or latency.  
  \item \textbf{Benchmark}: We release a systematic twenty-one-experiment matrix evaluated on LGG, BraTS 2020 and ISLES 2022, enabling reproducible study of scan-order effects in brain-MRI segmentation.  
  \item \textbf{Guidance}: Across all datasets, contiguous horizontal scans and one balanced four-direction order consistently achieve the highest Dice ranks, while the Friedman analysis confirms the statistical significance of scan choice. These simple orders deliver near-optimal accuracy with no compute penalty, giving practitioners an evidence-based shortlist of scans to try first when deploying Vision Mamba in clinical pipelines.  
\end{enumerate}

In short, our study establishes scan order as a consequential yet cost-free hyper-parameter for state-space vision models in neuro-imaging and provides both the tools and empirical evidence needed to exploit it.

\section{Related Work}

\subsection{State–Space Sequence Models}
State-Space Models (SSMs) are a cornerstone of control theory, providing a framework for modeling systems that evolve over time. A continuous-time linear SSM describes how an input signal $x(t)$ is mapped to an output signal $y(t)$ via a latent state vector $h(t) \in \mathbb{R}^N$:
\begin{equation}
  \dot{h}(t) = \mathbf{A}h(t) + \mathbf{B}x(t), \qquad
  y(t) = \mathbf{C}h(t) + \mathbf{D}x(t).
  \label{eq:ssm_continuous}
\end{equation}
The components of this system are:
\begin{itemize}
    \item $\mathbf{A} \in \mathbb{R}^{N \times N}$: The \textbf{state matrix}, which governs the internal dynamics of the system.
    \item $\mathbf{B} \in \mathbb{R}^{N \times D_{in}}$: The \textbf{input matrix}, which maps the input $x(t) \in \mathbb{R}^{D_{in}}$ to the state.
    \item $\mathbf{C} \in \mathbb{R}^{D_{out} \times N}$: The \textbf{output matrix}, which maps the state to the output $y(t) \in \mathbb{R}^{D_{out}}$.
    \item $\mathbf{D} \in \mathbb{R}^{D_{out} \times D_{in}}$: The \textbf{feedthrough matrix}, providing a direct skip connection from the input to the output.
\end{itemize}
For use with discrete sequential data, such as a sequence of image patches, the continuous system must be discretized. Using a fixed sampling step size $\Delta$, a common method like the zero-order hold transforms the continuous parameters $(\mathbf{A}, \mathbf{B})$ into discrete counterparts $(\bar{\mathbf{A}}, \bar{\mathbf{B}})$:
\begin{equation}
  h_{k} = \bar{\mathbf{A}}h_{k-1} + \bar{\mathbf{B}}x_{k}, \qquad
  y_{k} = \mathbf{C}h_{k} + \mathbf{D}x_{k},
  \label{eq:ssm_discrete}
\end{equation}
where $\bar{\mathbf{A}} = e^{\mathbf{A}\Delta}$ and $\bar{\mathbf{B}} = (\mathbf{A}\Delta)^{-1}(e^{\mathbf{A}\Delta} - \mathbf{I})\mathbf{B}\Delta$. The model can then be computed either recurrently as in Eq.~\ref{eq:ssm_discrete} or globally as a single long convolution.
The high computational cost of these operations historically limited their use in deep learning. The Structured State Space Sequence Model (S4) provided a breakthrough by constraining $\mathbf{A}$ to a specific diagonal-plus-low-rank structure, which enabled the use of highly efficient convolutional algorithms \cite{gu2021s4}. Mamba builds upon this by introducing an input-dependent selection mechanism that modulates the SSM parameters ($\mathbf{B}$, $\mathbf{C}$, and $\Delta$) for each input token. This selectivity, combined with a hardware-aware parallel scan algorithm, allows Mamba to match the performance of Transformers with linear time complexity \cite{gu2023mamba}.

\subsection{Sequence models in vision}

The Vision Transformer (ViT) attends across all patch tokens, but
storing a full attention matrix limits practical resolution.
Vision Mamba serialises patches and processes them with
bidirectional Mamba layers, achieving global context at linear cost
\cite{zhu2024vision}.  
VMamba inserts a four-direction selective-scan-2D (SS2D) block in a
hierarchical backbone \cite{liu2024vmamba}.  
PlainMamba explores serpentine scans that keep spatial neighbours
sequential \cite{yang2024plainmamba}.  
Zhu \textit{et al.} benchmarked twenty-two scan orders on aerial
imagery and found a left-to-right raster sufficient for land-cover
segmentation \cite{zhu2024rethinking}; whether that holds for brain
MRI remains open.

\subsection{Convolutional backbones for medical segmentation}

U-Net pairs an encoder–decoder with skip connections and remains the
cornerstone of biomedical segmentation \cite{ronneberger2015unet}.
Subsequent variants add dense links, residual paths, attention gates
or three-dimensional convolutions, but all convolutional designs still
aggregate context via stacked local filters. Surveys of deep medical
segmentation highlight the resulting limitations in modelling very
long-range dependencies \cite{kainz2024deep}.

\subsection{Brain-MRI segmentation benchmarks}

For focal brain-lesion segmentation the community relies chiefly on
BraTS 2020 for glioma tumours and ISLES 2022 for acute stroke.  The
lighter LGG dataset broadens tumour-grade diversity.  Together these
benchmarks cover the principal oncological and vascular pathologies
encountered in clinical neuroradiology \cite{hernandezpetzsche2022isles,menze2015brats,bakas2017advancing,bakas2018identifying, tcga_lgg}.

\subsection{Mamba variants for medical imaging and the open question of scan order}

Mamba blocks appear in several U-Net-like backbones.  U-Mamba inserts
one-dimensional SSM layers to extend the receptive field
\cite{ma2024umamba}.  
VM-UNet, MSVM-UNet and LKM-UNet replace convolutional encoders with
Vision-Mamba blocks and add multi-scale or large-kernel refinements
\cite{ruan2024vm,chen2024msvm,wang2024lkm}.  
SegMamba applies tri-orientated scans to three-dimensional volumes and
reports competitive results on the Medical Segmentation Decathlon
\cite{xing2024segmamba}.  
In almost every case the scan order is fixed to four orthogonal
passes; ablation is absent or restricted to a single dataset.

Brain MRI exhibits strong directional structure: diffusion-tensor
studies reveal anisotropic white-matter tracts aligned along preferred
axes \cite{alexander2007diffusion,lee2009anisotropic}.  Such priors
suggest that patch order could influence performance more on brain MRI
than on the irregular layouts of remote-sensing scenes.

\section{Methodology}

\subsection{Network Architecture}

\begin{figure}[t]
  \centering
  \includegraphics[width=\columnwidth]{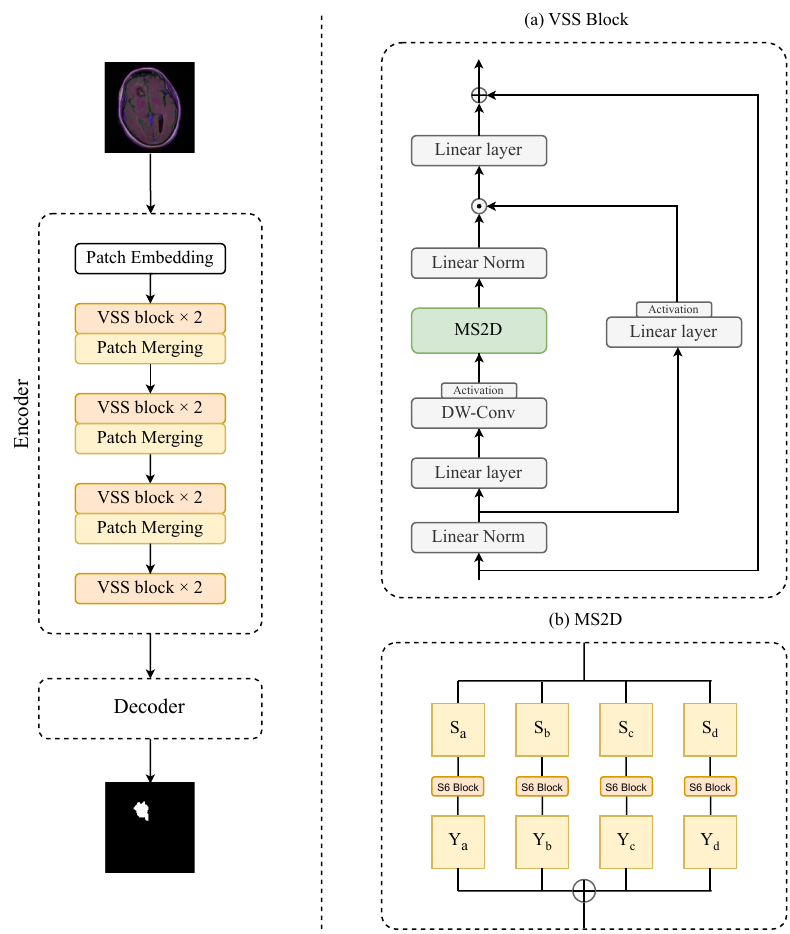}
  \caption{Overview of the proposed architecture.  
  \textbf{Left:} VM--UNet encoder–decoder with four resolution stages in the encoder and skip-connected, lightweight decoder.  
  \textbf{Right (a):} Internal structure of a VSS block, where the original SS2D unit is replaced by \textbf{MS2D}.  
  \textbf{Right (b):} \emph{MS2D} runs four parallel Mamba streams (\(S_a{\ldots}S_d\)); users supply 1, 2 or 4 unique scan IDs, which are duplicated so each block still evaluates four SSM kernels before summation.}
  \label{fig:architecture}
\end{figure}

Fig.~\ref{fig:architecture} presents an overview of the \textbf{VM\,-UNet} \cite{ruan2024vm} backbone and the proposed MS2D integration.  
VM--UNet is the first pure state–space \(U\)-shaped model for medical segmentation.  
After preprocessing, every axial slice has spatial size \(128\times128\) pixels and \(C_{\text{in}}\) modality channels,  
so its raw shape is \(128\times128\times C_{\text{in}}\).  
We divide the slice into non-overlapping \(4\times4\) patches, yielding a \(32\times32\) grid—i.e.\ \(32\times32 = 1024\) patch tokens—each token being a flattened vector of length \(4\times4\times C_{\text{in}}\).  
A shared \(1\times1\) linear projection maps every token to a fixed embedding dimension \(C = 96\) and adds a learnable two-dimensional positional bias.
The encoder comprises four resolution stages whose token grids are \((32\times32,\;16\times16,\;8\times8,\;4\times4)\).  
At every stage two residual \textit{VMamba} blocks are stacked.  
A block applies layer normalisation, a depth-wise \(3\times3\) convolution, the proposed \textit{MS2D} unit (Section \ref{sec:ms2d}) and a point-wise linear projection before residual addition.  
Between stages a \(2\times2\) strided convolution halves spatial size and doubles channel width, giving channel counts \((96,\;192,\;384,\;768)\).

The bottleneck at \(4\times4\) resolution adds two extra VMamba blocks.  
The decoder mirrors the encoder hierarchy with a lighter depth vector \([2,2,2,1]\).  
Each up-sampling step uses bilinear interpolation followed by a \(2\times2\) convolution and an element-wise skip connection from the matching encoder stage.  
A final \(1\times1\) convolution maps the \(96\)-channel tensor to a single foreground-logit map, followed by sigmoid activation.

Replacing SS2D with MS2D leaves the parameter count unchanged at \(27.43\,\text{M}\); floating-point operations are identical, so every scan experiment runs under equal capacity and compute.

\subsection{Multi Scan 2D (MS2D) Module}\label{sec:ms2d}

The original VMamba \textit{SS2D} exposes four fixed scan directions.  
To study patch order we replace SS2D with \textbf{MS2D}, which offers twelve primitive paths while keeping the workload constant.

\begin{figure*}[!t]
  \centering
  \includegraphics[width=\textwidth]{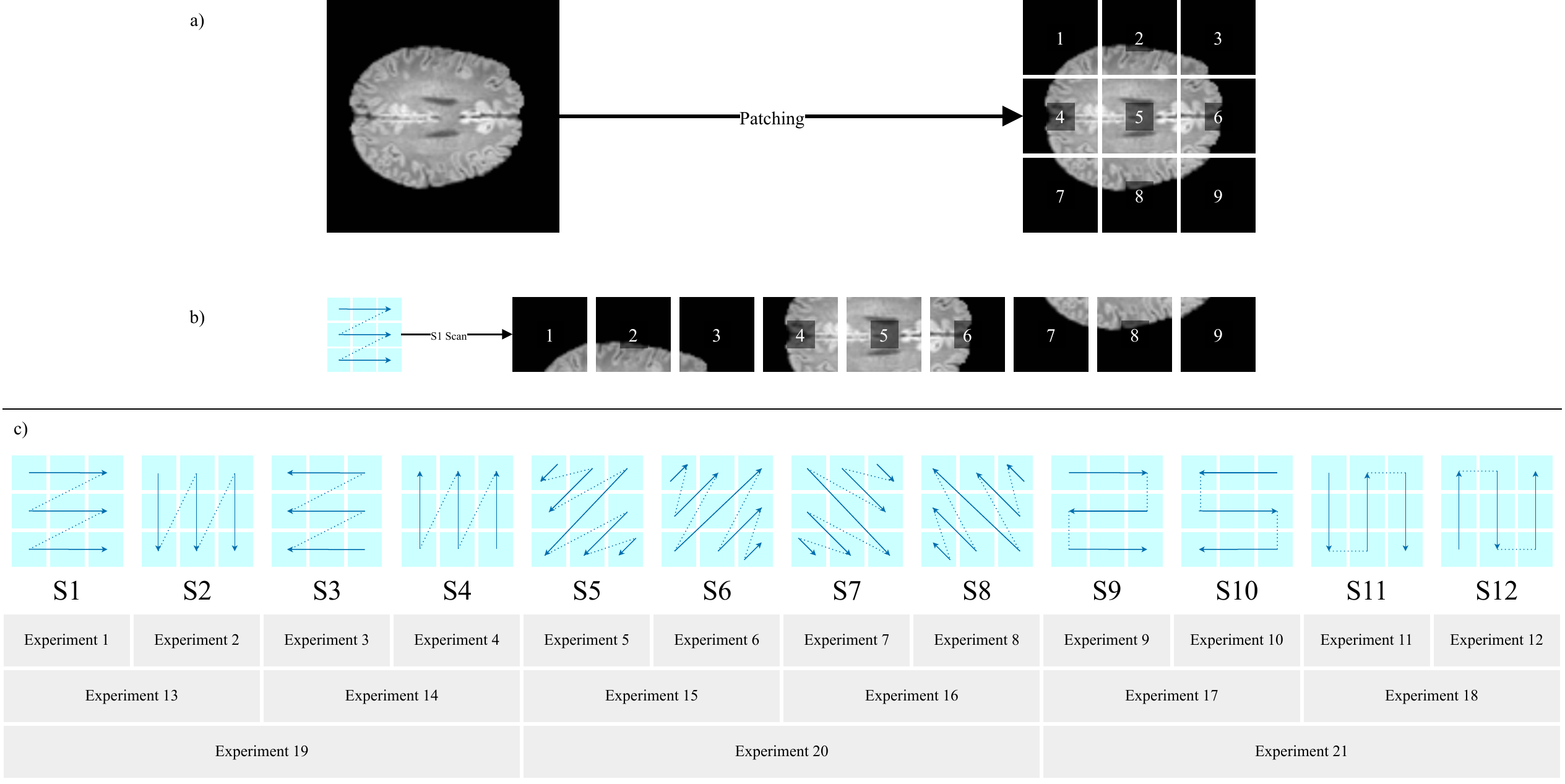}
  \caption{%
\textbf{Overview of scan permutations and experimental design.}
\emph{(a)}~An axial ISLES slice is patched into a \(3\times3\) grid; numbers mark the visitation order for the chosen scan.  
\emph{(b)}~Example sequence produced by the single left-to-right raster (S1); the remaining primitive scans S2–S12 generate analogous sequences and are omitted for space.  
\emph{(c)}~Experiment matrix.  The top row depicts the 12 primitive scans (S1–S12).  
Experiments 1–12 evaluate each scan in isolation (\(k{=}1\));  
Experiments 13–18 pair two paths (\(k{=}2\));  
Experiments 19–21 aggregate four paths (\(k{=}4\)).  
}
  \label{fig:scan_paths}
\end{figure*}

\textit{Scan bank}: Twelve paths are defined: two row rasters (left-to-right, right-to-left), two column rasters (top-to-bottom, bottom-to-top), four main diagonals and four serpentine traversals.  Each path is a permutation \(\pi_d\) of the patch grid.

\textit{Stream selection}: MS2D always executes four parallel VMamba streams.  
Users choose \(k\in\{1,2,4\}\) unique scan IDs:
\begin{itemize}
  \item \(k=1\): the chosen path is copied four times;
  \item \(k=2\): each path is copied twice;
  \item \(k=4\): every path is used once.
\end{itemize}
This duplication rule guarantees identical FLOPs in every experiment.

\textit{Processing}: For stream \(i\) the feature map \(\mathbf V\in\mathbb R^{H\times W\times C}\) is rearranged into sequence form \(\mathbf X_i\) by \(\pi_{d_i}\).  
After layer normalization, each sequence passes through an SSM–Mamba layer,
\[
  \mathbf Y_i = \mathrm{Mamba}\bigl(\mathbf X_i;\,\theta_i\bigr)\in\mathbb R^{L\times C}.
\]
The four outputs are reshaped to spatial tensors and summed element-wise,
\[
  \mathbf Z=\sum_{i=1}^{4}\mathbf Y_i.
\]
A \(1\times1\) linear layer mixes channels and the result joins the residual path of the VMamba block.

\textit{Complexity}: Summation preserves channel width, so MS2D adds no parameters.  
Each block still performs four SSM convolutions, yielding the same \(\mathcal O(LC)\) time and \(\mathcal O(C)\) memory as the baseline.

\subsection{Primitive Scan Catalogue}
Fig.~\ref{fig:scan_paths} visualizes the twelve paths (S1–S12); each produces a permutation \(\pi_d\) that maps grid coordinates \((r,c)\) to a one-dimensional index \(t\).

\subsection{Experiment Matrix}

To isolate the \emph{patch-order} effect we ran a total of \textbf{21 experiments}, each defined by the set of scan directions fed into the Multi-Scan 2D (MS2D) block.  Fig. \ref{fig:scan_paths}\,(c) shows every scan pattern (blue grids, labelled \textsf{S1–S12}) and maps them to the corresponding experiment numbers listed below.

\paragraph*{Single–direction experiments (1–12).}  
Experiments 1–12 each use \emph{one} of the twelve primitive scan paths (\textsf{S1–S12}).  
Because MS2D always ingests four streams, the chosen path is duplicated four times \((k=1)\).  
For example, Experiment 5 repeats the first diagonal scan \textsf{S5} four times.

\paragraph*{Bidirectional experiments (13–18).}  
Experiments 13–18 combine \emph{two complementary} directions, each supplied twice \((k=2)\):

\begin{itemize}
  \item \textbf{Experiment 13:} \textsf{S1} (left $\rightarrow$ right) + \textsf{S2} (top $\rightarrow$ bottom)  
  \item \textbf{Experiment 14:} \textsf{S3} (right $\rightarrow$ left) + \textsf{S4} (bottom $\rightarrow$ top)  
  \item \textbf{Experiment 15:} \textsf{S5} + \textsf{S6} (diagonals)  
  \item \textbf{Experiment 16:} \textsf{S7} + \textsf{S8} (opposite diagonals)  
  \item \textbf{Experiment 17:} \textsf{S9} + \textsf{S10} (horizontal serpentines in opposite order)  
  \item \textbf{Experiment 18:} \textsf{S11} + \textsf{S12} (vertical serpentines in opposite order)
\end{itemize}

\paragraph*{Four-direction mixes (19–21).}  
Experiments 19–21 feed \emph{four distinct} directions once each \((k=4)\), thereby exercising the full capacity of MS2D without duplication:

\begin{itemize}
  \item \textbf{Experiment 19 (Horizontal + Vertical):} \textsf{S1, S2, S3, S4}  
  \item \textbf{Experiment 20 (All diagonals):} \textsf{S5, S6, S7, S8}  
  \item \textbf{Experiment 21 (All serpentines):} \textsf{S9, S10, S11, S12}
\end{itemize}

Across all 21 experiments every MS2D invocation still comprises exactly four Mamba kernels, and we keep \emph{all other} training hyper-parameters, loss functions and data-augmentation settings identical.  Consequently, any performance variation reported in Section \ref{sec:results} can be attributed solely to the ordering of patches explored in Figure \ref{fig:scan_paths}\,(c).

\section{Experimental Setup}

\begin{table*}[!t]
  \caption{Brain-MRI segmentation datasets used in this study}
  \label{tab:datasets}
  \centering
  \renewcommand{\arraystretch}{2}   
  \begin{tabular}{@{}l l p{5.5cm} c c c@{}}
    \hline
    \textbf{Dataset} & \textbf{Input modality} & \textbf{Target classes} & \textbf{Cases} & \textbf{\#Slices} & \textbf{Task} \\[-6pt]
    &&& (train / val / test) && \\
    \hline
    ISLES 2022 & DWI & Stroke core vs. background & 175 / 38 / 37 & 15\,684 & Binary \\
    BraTS 2020 & T1, T1-c, T2, FLAIR &Background, NCR/NET, ED, ET & 259 / 55 / 55 & 57\,195 & 3-class \\
    LGG-MRI & T2-w, FLAIR, T1-c (RGB stack) & Tumour vs. background & 77 / 17 / 16 & 4\,976 & Binary \\
    \hline
  \end{tabular}
\end{table*}

To ensure a fair and reproducible comparison across all 21 scan strategies, we established a rigorous and standardized experimental protocol. All models were trained and evaluated using identical data splits, preprocessing pipelines, augmentation strategies, and training configurations.
\subsection{Datasets and Preprocessing}
We utilized three publicly available brain MRI datasets, as detailed in Table~\ref{tab:datasets}, each targeting a distinct pathology.
\begin{itemize}
    \item \textbf{Data Splitting:} For each dataset, we performed a subject-level split, assigning 70\% of cases to the training set, 15\% to validation, and 15\% to the test set. This strict separation at the case level is crucial to prevent data leakage and ensure that the model is evaluated on entirely unseen patient data.
    \item \textbf{Preprocessing Pipeline:} All MRI volumes underwent a standardized preprocessing pipeline before being fed to the network. First, we applied \emph{min–max normalization}, scaling each volume’s intensities to the \([0,1]\) range by subtracting the minimum and dividing by the range of the non-zero voxels. This harmonizes intensity distributions across subjects while preserving relative contrast. Subsequently, each axial slice was resized to a uniform spatial resolution of \(128 \times 128\) pixels to ensure consistent input dimensions for the model.

\end{itemize}

\subsection{Training Protocol}
All experiments were conducted with a consistent training regime to isolate the effect of the scan order.
\begin{itemize}
    \item \textbf{Optimizer:} We used the \texttt{AdamW} optimizer with $\beta_{1}=0.9$ and $\beta_{2}=0.999$. \texttt{AdamW} is preferred over the standard \texttt{Adam} for transformer-based models as its decoupled weight decay ($10^{-2}$) regularizes the model more effectively, often leading to better generalization.
    \item \textbf{Learning Rate Schedule:} The learning rate was managed by a \texttt{CosineAnnealingLR} schedule, starting at an initial rate of $\mathrm{LR}_{0}=10^{-3}$ and gradually annealing to a minimum of $\mathrm{LR}_{\min}=10^{-6}$ over the course of training.
    \item \textbf{Batch Size:} Models were trained with a batch size of 48.
    \item \textbf{Best Model:} The model weights corresponding to the lowest validation loss were saved for testing.
\end{itemize}
\subsection{Loss Functions}
Our choice of loss function was tailored to the specific segmentation task of each dataset.
\begin{itemize}
    \item \textbf{Binary Segmentation (ISLES, LGG):} We used a composite loss, $\mathcal{L}_{\mathrm{BCE+Dice}} = \mathcal{L}_{\mathrm{BCE}} + \mathcal{L}_{\mathrm{Dice}}$. This combination is highly effective for medical imaging; the Binary Cross-Entropy term ($\mathcal{L}_{\mathrm{BCE}}$) promotes pixel-level correctness, while the Dice loss ($\mathcal{L}_{\mathrm{Dice}}$) directly optimizes the geometric overlap between the prediction and the ground truth, which is particularly robust to class imbalance.
    \item \textbf{Multi-class Segmentation (BraTS):} For this more complex task, we used a weighted sum of Cross-Entropy and multi-class Dice loss: $\mathcal{L}_{\mathrm{CE+mDice}} = 0.4 \cdot \mathcal{L}_{\mathrm{CE}} + 0.6 \cdot \mathcal{L}_{\mathrm{mDice}}$. Giving a slightly higher weight to the multi-class Dice term ($\mathcal{L}_{\mathrm{mDice}}$) helps the model better handle the severe class imbalance between the different tumor sub-regions (e.g., necrotic core, edema, enhancing tumor).
\end{itemize}
\subsection{Evaluation and Hardware}
Model performance was primarily evaluated using the Dice Similarity Coefficient (DSC) for its sensitivity to overlap and the mean Intersection-over-Union (mIoU) for a complementary perspective on segmentation accuracy. The statistical significance of performance differences was assessed using the Friedman test, as described in Section V-C. All experiments were conducted on an Azure cloud instance equipped with two NVIDIA V100 GPUs (16 GB).

\section{Results}\label{sec:results}

\begin{table*}[!t]
  \caption{Test-set Dice (DSC) and mIoU for the 21 scan strategies.
           Best Dice per dataset is highlighted in \textbf{bold}.}
  \label{tab:test_metrics}
  \centering
  \footnotesize
  \setlength{\tabcolsep}{25pt}
  \renewcommand{\arraystretch}{1.15}
  \begin{tabular}{@{}lc cc cc cc@{}}
    \toprule
    \multirow{2}{*}{\textbf{Experiment}} &
      \multicolumn{2}{c}{\textbf{BraTS 2020}} &
      \multicolumn{2}{c}{\textbf{ISLES 2022}} &
      \multicolumn{2}{c}{\textbf{LGG-MRI}} \\
    \cmidrule(lr){2-3}\cmidrule(lr){4-5}\cmidrule(lr){6-7}
      & Dice & mIoU & Dice & mIoU & Dice & mIoU \\ \midrule
Exp 1 & 0.739 & 0.588 & 0.643 & 0.474 & 0.674 & 0.508 \\
Exp 2 & 0.721 & 0.569 & 0.811 & 0.682 & 0.697 & 0.535 \\
Exp 3 & \textbf{0.753} & 0.607 & 0.815 & 0.687 & 0.740 & 0.587 \\
Exp 4 & 0.701 & 0.544 & 0.767 & 0.622 & 0.702 & 0.541 \\
Exp 5 & 0.686 & 0.528 & 0.740 & 0.588 & 0.642 & 0.473 \\
Exp 6 & 0.639 & 0.476 & 0.710 & 0.551 & 0.646 & 0.477 \\
Exp 7 & 0.688 & 0.534 & 0.551 & 0.380 & 0.629 & 0.459 \\
Exp 8 & 0.701 & 0.548 & 0.666 & 0.499 & 0.692 & 0.529 \\
Exp 9 & 0.689 & 0.529 & 0.588 & 0.416 & 0.637 & 0.468 \\
Exp 10 & 0.729 & 0.577 & 0.636 & 0.466 & 0.647 & 0.478 \\
Exp 11 & 0.718 & 0.565 & 0.762 & 0.615 & 0.728 & 0.573 \\
Exp 12 & 0.716 & 0.563 & 0.781 & 0.641 & 0.711 & 0.552 \\
Exp 13 & 0.745 & 0.597 & 0.757 & 0.610 & 0.727 & 0.572 \\
Exp 14 & 0.695 & 0.537 & 0.815 & 0.688 & 0.723 & 0.566 \\
Exp 15 & 0.705 & 0.553 & 0.718 & 0.560 & 0.661 & 0.493 \\
Exp 16 & 0.665 & 0.507 & 0.581 & 0.409 & 0.648 & 0.479 \\
Exp 17 & 0.722 & 0.568 & 0.579 & 0.407 & 0.624 & 0.454 \\
Exp 18 & 0.728 & 0.576 & 0.769 & 0.625 & 0.720 & 0.563 \\
Exp 19 & 0.731 & 0.581 & \textbf{0.820} & 0.694 & \textbf{0.746} & 0.595 \\
Exp 20 & 0.673 & 0.515 & 0.754 & 0.605 & 0.666 & 0.500 \\
Exp 21 & 0.734 & 0.584 & 0.801 & 0.667 & 0.705 & 0.545 \\ \bottomrule
  \end{tabular}
\end{table*}

Our experiments reveal that the choice of patch scanning order is not a trivial implementation detail but a critical hyperparameter with a statistically significant and practically meaningful impact on segmentation performance. This section details the quantitative outcomes, highlights the performance disparities between scan strategies, and presents the statistical analysis that confirms these observations.

\subsection{Overall Performance and Impact of Scan Strategy}
\begin{figure*}[!t]
  \centering
  \includegraphics[width=\textwidth]{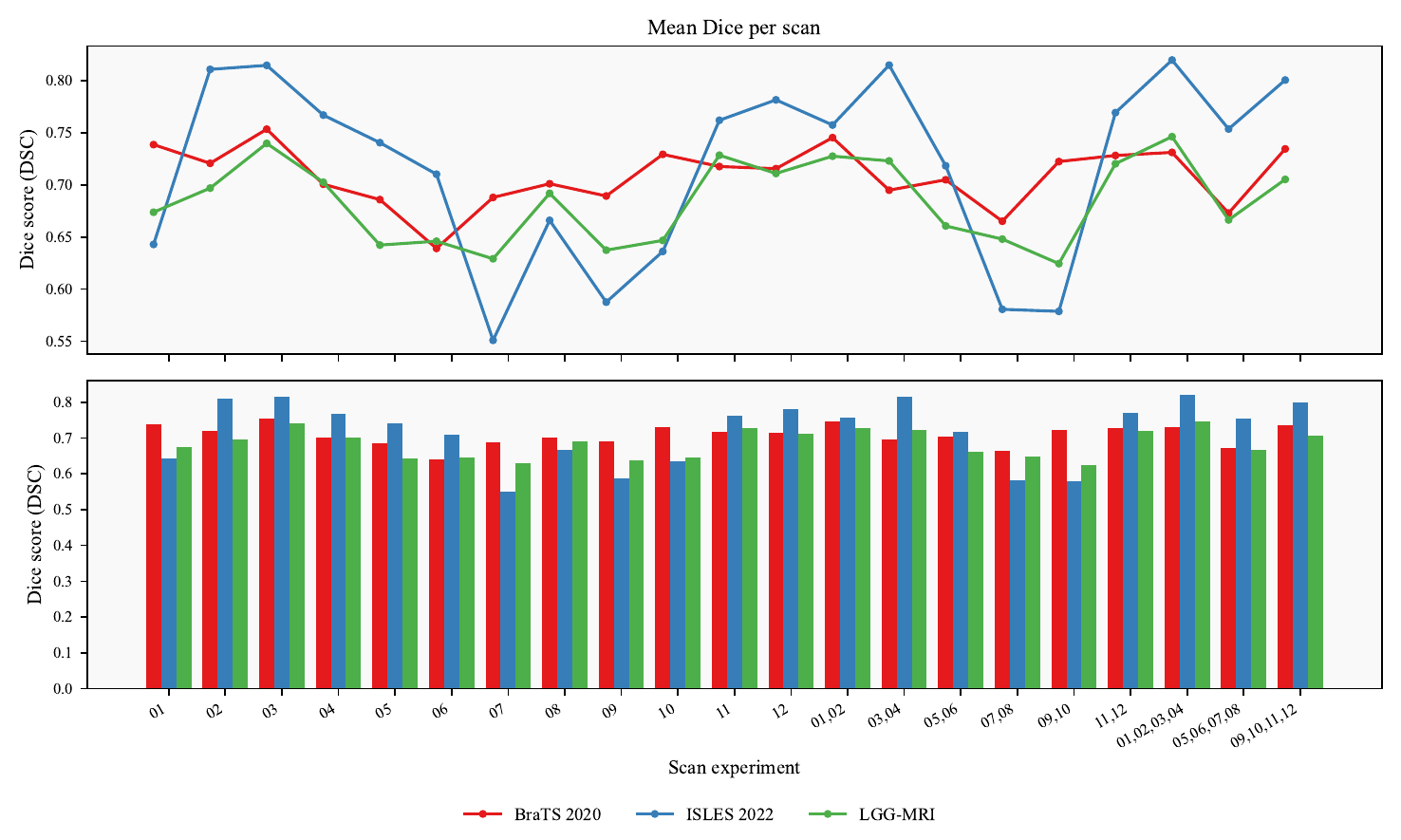}%
  \caption{Mean Dice scores of the 21 scan strategies on BraTS 2020,
           ISLES 2022, and LGG-MRI.  Top: line plot; bottom: bar chart.}
  \label{fig:dice_vis}
\end{figure*}

The complete test-set results for all 21 scan strategies across the three datasets are presented in Table~\ref{tab:test_metrics}, with a visual summary provided in Figure~\ref{fig:dice_vis}. The data reveals a wide performance variance directly attributable to the scan order. The delta between the best and worst-performing strategies is substantial across all datasets, with Dice score ranges of 0.114 on BraTS 2020 (from 0.639 to 0.753), 0.122 on LGG-MRI (from 0.624 to 0.746), and a remarkable 0.269 on ISLES 2022 (from 0.551 to 0.820). This demonstrates that a simple, almost cost-free change in scan path can lead to dramatic gains or losses in segmentation accuracy.

\subsection{Identifying the Most Effective Scan Paths}
A clear pattern of superior performance emerges for contiguous, linear scan paths.
\begin{itemize}
    \item The single \textbf{right-to-left horizontal scan (Exp. 3)} proved to be a consistently strong performer, achieving the highest Dice score on the BraTS 2020 dataset (0.753) and ranking second on both ISLES 2022 (0.815) and LGG-MRI (0.740).
    \item The four-direction mix of \textbf{horizontal and vertical scans (Exp. 19)} also demonstrated exceptional and robust performance. It secured the top Dice scores for both the ISLES 2022 (0.820) and LGG-MRI (0.746) datasets, underscoring the benefit of combining orthogonal perspectives.
    \item Bidirectional horizontal scans (\textbf{Exp. 13}) and vertical serpentine scans (\textbf{Exp. 18}) also consistently placed in the upper tier of performers, reinforcing the advantage of spatially coherent, raster-like traversals.
\end{itemize}

\subsection{Identifying the Least Effective Scan Paths}
Conversely, scan strategies that disrupt spatial locality, such as diagonal or complex serpentine paths, consistently underperformed.
\begin{itemize}
    \item The \textbf{diagonal S7} was unequivocally the worst-performing strategy, yielding the lowest Dice score on two of the three datasets, including a particularly low 0.551 on ISLES 2022. Its bidirectional counterpart, \textbf{Exp. 16}, was also among the poorest performers.
    \item Other diagonal traversals (\textbf{S5, S6, S8}) and the horizontal serpentine scan (\textbf{S9}) also populated the bottom ranks, suggesting that their non-contiguous patch sequences hinder the model's ability to learn effective spatial representations for brain MRI.
\end{itemize}

\subsection{Statistical Validation of Scan Order's Significance}
A Friedman test on the Dice-score matrix (three datasets $\times$ 21 scan strategies) yielded $\chi^{2}_{20}=43.86$ and a p-value of 0.0016.  
Because $p<0.05$, we reject the null hypothesis and conclude that scan order has a statistically significant impact on segmentation performance.

The Friedman mean-rank ordering places \textbf{Exp. 3} first (mean rank 2.00), closely followed by the orthogonal four-direction mix \textbf{Exp. 19} (2.33). At the opposite end, the diagonal paths \textbf{Exp. 7} (19.33) and \textbf{Exp. 16} (18.00) rank last. These results reinforce the empirical finding that simple, spatially contiguous raster scans consistently outperform disjointed diagonal traversals for Mamba-based brain-MRI segmentation.

\section{Discussion}

The results compellingly demonstrate that patch scan order is a determinant of performance for Mamba-based MRI segmentation. Our statistical analysis not only confirms that the effect is significant ($\chi^{2}_{20}=43.9, p=0.0016$) but the performance deltas—up to a 27-point Dice swing—highlight its practical importance. This discussion dissects these findings, offering explanations for the observed performance hierarchy and outlining the implications for future research.

\subsection{The Primacy of Contiguous Scans for Spatial Coherence}
The superior performance of simple, contiguous raster scans (horizontal and vertical) is the study's central finding. This outcome can be attributed to the inherent nature of State Space Models (SSMs) like Mamba, which process data as one-dimensional sequences. For an SSM to effectively model a 2D structure, the 1D sequence must retain as much spatial locality as possible.

Brain pathologies such as tumors and strokes are typically contiguous masses. A simple raster scan (e.g., S3) traverses these regions coherently, presenting adjacent image patches as adjacent tokens in the sequence. This allows the Mamba model to efficiently propagate context and build a representation of the lesion's shape and texture along a continuous path. The success of the four-direction orthogonal mix (Exp. 19) further supports this, suggesting that providing the model with multiple, distinct, and spatially coherent "views" of the image allows it to form a more robust and comprehensive understanding of 2D spatial dependencies.

\subsection{The Inefficiency of Disjointed and Redundant Scan Paths}
In contrast, the poorest-performing strategies were diagonal and complex serpentine scans. Their failure likely stems from the disruption of spatial locality. A diagonal scan, for instance, creates large spatial "jumps" between consecutive tokens in the sequence forcing the model to expend its limited capacity on bridging these artificial spatial gaps rather than on semantic feature extraction.

Furthermore, the analysis of combined scans suggests that more is not always better. Bidirectional pairs (e.g., Exp. 13, left-to-right and top-to-bottom) rarely outperformed the best single direction within the pair. The most effective combinations are those that are complementary, not repetitive, such as the fusion of orthogonal scans.

\subsection{Limitations and Avenues for Future Research}
Our study, while comprehensive, is bounded by certain limitations which naturally illuminate avenues for future work.
\begin{itemize}
    \item \textbf{Scope of Generalization:} The experiments were conducted on three brain MRI datasets and a single, albeit representative, VMamba-based architecture. Future work should validate these findings across a broader range of medical imaging tasks (e.g., multiple sclerosis lesion segmentation), modalities (e.g., CT, PET), and other Mamba-inspired vision backbones to determine if the preference for contiguous scans is a universal principle.

    \item \textbf{Fixed Hyperparameters:} All experiments were run with a fixed image and patch size. The interaction between patch resolution and optimal scan order remains an open question. Larger patches inherently average more local information, which might alter the impact of the scan path.

    \item \textbf{Static vs. Dynamic Scanning:} This work evaluated a fixed set of predefined scan strategies. A significant next step would be to develop models with \textbf{adaptive scan scheduling}. A lightweight, learnable module could be designed to dynamically select the most appropriate scan path(s) on a per-image or even per-region basis, potentially tailoring the information flow to the specific anatomical structures present in a slice.

    \item \textbf{Evaluation Protocol:} While a standard train-validation-test split was used, employing a more rigorous k-fold cross-validation protocol in future studies would provide tighter confidence intervals on performance metrics and further verify the robustness of the scan order rankings.
\end{itemize}

\section{Conclusion}

We have conducted the first systematic study of \emph{scan-order effects} in state-space vision models for brain-MRI segmentation. By introducing the parameter-neutral \textbf{Multi-Scan-2D (MS2D)} module into a VM-UNet architecture, we benchmarked 21 distinct scan strategies across three public datasets covering diverse brain pathologies. The results are unequivocal: the method of serializing a 2D image into a 1D sequence is not a benign implementation detail but a critical factor for model performance. A non-parametric Friedman test confirms this with high statistical significance ($\chi^{2}_{20}=43.9, p=0.0016$), while the observed performance variance—with Dice score swings of up to 27 points—highlights the profound practical implications.

Our findings translate into direct, actionable guidance for researchers and practitioners:

\begin{enumerate}
    \item \textbf{Prioritize Spatially Coherent Scans.} The most effective strategies are simple, contiguous paths. Standard horizontal and vertical raster scans (e.g., S3) and their orthogonal combination (Exp. 19) consistently deliver top-tier accuracy. These paths preserve the spatial locality of anatomical features, allowing the sequence model to build context effectively.

    \item \textbf{Avoid Spatially Disjointed Paths.} Diagonal and complex serpentine scans consistently yield the poorest results. By creating artificial gaps between adjacent tokens in the sequence, these paths disrupt the model's ability to learn local relationships, hindering its performance.

    \item \textbf{Treat Scan Order as a Free Hyperparameter.} Because the choice of scan path can be altered without affecting model parameters or computational cost, it should be treated as a primary hyperparameter. Practitioners should screen a shortlist of promising contiguous scans early in the development pipeline to maximize model efficacy with no additional overhead.
\end{enumerate}

By systematically demonstrating the impact of patch ordering and identifying a set of evidence-based best practices, this work moves the field beyond fixed, arbitrary scanning heuristics. We provide both the tools and the rationale needed to make informed decisions about this crucial design choice, paving the way for more robust, efficient, and ultimately more reliable deployment of state-space models in clinical and research environments.
\section*{Acknowledgment}
We gratefully acknowledge the financial support provided by the Qatar National Research Fund (QNRF), a member of the Qatar Research, Development, and Innovation (QRDI) Council, through the Undergraduate Research Experience Program (UREP) under grant number UREP31-077-2-025.


\bibliographystyle{IEEEtran}
\bibliography{refs}

\end{document}